%
\def\ptitlea{Construction of exact solutions to eigenvalue problems}
\def\ptitleb{by the asymptotic iteration method}
\font\tr=cmr10                          
\font\bf=cmbx10                         
\font\it=cmti10                         
\font\trbig=cmbx10 scaled 1500          
\font\tiny=cmr8                         
\def\title#1{\bigskip\noindent\bf #1 ~ \tr\smallskip} 
\output={\shipout\vbox{\makeheadline
                                      \ifnum\the\pageno>1 {\hrule}  \fi 
                                      {\pagebody}   
                                      \makefootline}
                   \advancepageno}

\headline{\noindent {\ifnum\the\pageno>1 
                                   {\tiny \ptitle\hfil
page~\the\pageno}\fi}}
\footline{}
\tr 
\def\nl{\hfil\break\noindent}  
\def\ni{\noindent}             
\def\np{\hfil\vfil\break}

\def\hi#1#2{$#1$\kern -2pt-#2} 
\def\hy#1#2{#1-\kern -2pt$#2$} 
 

\baselineskip 15 true pt  
\parskip=0pt plus 5pt 
\parindent 0.25in
\hsize 6.0 true in 
\hoffset 0.25 true in 
\emergencystretch=0.6 in                 
\vfuzz 0.4 in                            
\hfuzz  0.4 in                           
\vglue 0.1true in
\mathsurround=2pt                        
\topskip=24pt                            

\output={\shipout\vbox{\makeheadline
                                      \ifnum\the\pageno>1 {\hrule}  \fi
                                      {\pagebody}
                                      \makefootline}
                   \advancepageno}
\headline{\noindent {\ifnum\the\pageno>1
                                   {\tiny \ptitlea\hfil
page~\the\pageno}\fi}}
\footline{}
\def\nl{\hfil\break\noindent}  
\def\ni{\noindent}             
\def\np{\hfil\vfil\break}
\def\ppl#1{{\noindent\leftskip 11 cm #1\vskip 0 pt}} 

\def\hi#1#2{$#1$\kern -2pt-#2} 
\def\hy#1#2{#1-\kern -2pt$#2$} 


\def\frac#1#2{{{#1}\over{#2}}}
\newcount\zz  \zz=0  
\newcount\q   
\newcount\qq    \qq=0  
\def\pref#1#2#3#4#5{\frenchspacing \global \advance \q by 1     
    \edef#1{\the\q}{\ifnum \zz=1{\item{{[\the\q]}}{#2}{\bf #3},{ #4.}{~#5}\medskip} \fi}}
\def\bref #1#2#3#4#5{\frenchspacing \global \advance \q by 1     
    \edef#1{\the\q}
    {\ifnum \zz=1 { %
       \item{{[\the\q]}}
       {#2}, {\it #3} {(#4).}{~#5}\medskip} \fi}}
\def\gref #1#2{\frenchspacing \global \advance \q by 1  
    \edef#1{\the\q}
    {\ifnum \zz=1 { %
       \item{{[\the\q]}}
       {#2.}\medskip} \fi}}

 \def\sref #1{\kern 3pt[#1]}
 \def\srefs#1#2{{\kern 3pt[#1-#2]}}
\def\references#1{\zz=#1
   \parskip=2pt plus 1pt   
   {\ifnum \zz=1 {\noindent \bf References \medskip} \fi} \q=\qq
\pref{\lev} {G. L\`evai, J. Phys. A: Math. Gen. }{25}{L521 (1992)}{}
\pref{\gen} {L. Gendenshtein, Zh. Eksp. Teor. Fiz. Pis. Red. }{38}{299 (1983)}{Engl. transl. JETP lett. {\bf 38} 356 (1983)}
\bref{\dirac}{P. A. M. Dirac}{Quantum Mechanics}{Clarendon Press, Oxford, 1930}{}
\pref{\infh}{L. Infeld and T. E. Hull, Rev. Mod. Phys.}{23}{21 (1951)}{}
\pref{\stahl}{A. Stahlhofen, Il Nuovo Cimento B}{104}{447 (1989)}{}
\bref{\delang}{O. L. de Lange and R. E. Raab}{Operator Methods in Quantum Mechanics}
{Clarendon Press, Oxford, 1991}{}
\pref{\cks}{F. Cooper, A. Khare and U. Sukhatme, Phys. Rept. }{215}{267 (1995)}{}
\pref{\lmb}{L. M. Berkovich, Proc. Inst. Math. NAS Ukraine }{30}{25 (2000)}{}
\pref{\erm}{R. M. Edelstein, K. S. Govinder and F. M. Mahomed, J. Phys. A }{34}{1141 (2001)}{}
\pref{\lkp}{G. L\`evai, B. K\`onya, and Z. Papp, J. Math. Phys. }{39}{5811 (1998)}{}
\pref{\mp}{D. A. Morales and Z. Parra-Mejías,  Canad. J. Phys. }{77}{ 863 (1999)}{}
\pref{\cm}{R. N. Chaudhuri and M. Mondal, Phys. Rev. A }{52}{ 1850 (1995)}{}
\pref{\chs}{H. Ciftci, R. L. Hall and N. Saad, J. Phys. A: Math. Gen. }{36}{11807 (2003)}{}
\pref{\bw}{C. M. Bender and Q. Wang, J. Phys. A }{34}{9835 (2001)}{}
\pref{\fern}{F. M. Fernandez, J. Phys. A }{37}{6173 (2004)}{}
\pref{\pt}{G. P\"oschl and E. Teller, Z. Physik }{83}{143 (1933)}{}
\pref{\rmp}{N. Rosen and P. M. Morse, Phys. Rev. }{42}{210 (1932)}{}
\pref{\lot}{W. Lotmar, Z. Physik }{93}{528 (1935)}{}
\bref{\aar}{G. E. Andrews, R. Askey, and R. Roy}{Special Functions}{Cambridge University Press, 1999}{ Definition 2.5.1 p 99}
\pref{\nie}{M. M. Nieto, Phys. Rev. A }{17}{ 1273 (1978)}{}
}
\references{0}    
\ppl{CUQM-107}
\ppl{math-ph/0412030}
\ppl{December 2004}
\vskip 1.0 true in
\centerline{\bf\trbig\ptitlea}
\vskip 0.2 true in
\centerline{\bf\trbig\ptitleb}
\medskip
\vskip 0.25 true in
\centerline{Hakan Ciftci$^\dagger$$^*$, Richard L. Hall$^\dagger$ and Nasser Saad$^\ddagger$}
\vskip 0.25 true in
{\leftskip=0pt plus 1fil
\rightskip=0pt plus 1fil\parfillskip=0pt\baselineskip 10 pt
\obeylines
$^\dagger$ Department of Mathematics and Statistics, Concordia University
1455 de Maisonneuve Boulevard West, Montr\'eal
Qu\'ebec, Canada H3G 1M8.
E-mail: rhall@mathstat.concordia.ca\par}
\medskip
{\leftskip=0pt plus 1fil
\rightskip=0pt plus 1fil\parfillskip=0pt\baselineskip 10 pt
\obeylines
$^\ddagger$ Department of Mathematics and Statistics
University of Prince Edward Island
550 University Avenue, Charlottetown
PEI, Canada C1A 4P3.
E-mail: nsaad@upei.ca\par}
\medskip
{\leftskip=0pt plus 1fil \rightskip=0pt plus 1fil\parfillskip=0pt
\baselineskip 10 pt
\obeylines $^*$ Gazi Universitesi, Fen-Edebiyat Fak\"ultesi 
Fizik B\"ol\"um\"u, 06500 Teknikokullar 
Ankara, Turkey.
E-mail: hciftci@mathstat.concordia.ca and hciftci@gazi.edu.tr\par}
\vskip 0.5 true in
\centerline{\bf Abstract}\bigskip\noindent We apply the asymptotic iteration method (AIM) 
 [J. Phys. A: Math. Gen. {\bf 36}, 11807 (2003)] to
solve new classes of second-order homogeneous linear differential equation. In particular, solutions are found for a general class of eigenvalue problems which includes  Schr\"odinger problems with Coulomb, harmonic oscillator, or P\"oschl-Teller potentials, as well as the special eigenproblems studied recently by Bender {\it et al}~[J. Phys. A: Math. Gen. {\bf 34} 9835 (2001)]  and generalized in the present paper to higher dimensions.

\vskip 0.5 true in

\vfil\eject
\title{1. Introduction}
\medskip
\noindent The study of exactly solvable problems in quantum mechanics and the relation between their solutions represent valuable contributions to mathematical physics.  The range of potentials for which Schr\"odinger's equation can be solved exactly has been extended considerably owing to the investigations inspired, for example, by super-symmetric quantum mechanics\sref{\lev}, shape invariance\sref{\gen}, and the factorization method\sref{\dirac-\erm}. Coulomb, harmonic oscillator, and P\"oschl-Teller potentials are known examples of exactly solvable problems. It is also known that Coulomb and harmonic oscillator potentials are inter-related and that they are members of a family of exactly solvable Schr\"odinger equations\sref{\lkp-\cm}. In this article, we use the asymptotic iteration method (AIM)\sref{\chs} to demonstrate that P\"oschl-Teller potentials also belong to this family and that the (Coulomb, harmonic oscillator, P\"oschl-Teller) exact results may be obtained by elementary transformations, as particular cases of the following second-order homogeneous linear differential equation:
$$y^{\prime\prime}=2\left({{a~x^{N+1}}\over{1-bx^{N+2}}}-{{(m+1)}\over{x}}\right)y^\prime-{{wx^N}\over{1-bx^{N+2}}}\ y,\eqno(1.1)$$
\nl where $0\leq x< b^{1\over N+2}$ if $b>0$ and $0\leq x< \infty$ if $b\leq 0$.
Here $N=-1,0,1,\dots,$ and the real numbers  $a$, $m$, and $w$ are to be specified later.\medskip 

We should like to make clear at the outset that all the problems discussed in this paper can be transformed into special cases of (1.1).  The discovery of a common source for the entire set of problems and their treatment in arbitrary dimension $d$ by the asymptotic iteration method (AIM) are the most significant features of the paper. Our work has two related aspects: the solution of differential equations, and the solution of boundary-value problems.  In particular we shall show that exact solutions of the class of eigenvalue problem of the form 
$$-y^{\prime\prime}(x)+x^{2N+2}y(x)=E x^{N}y(x)\quad\ (N=-1,0,1,\dots)\eqno(1.2)$$
recently studied by Bender {\it et al}\sref{\bw} follow as special cases of the same differential equation (1.1). This will allow us to extend the solutions of (1.2) to provide analytic formulae for the solutions of the eigenvalue problem  
$$-y^{\prime\prime}(x)+\bigg({m(m+1)\over x^2}+x^{2N+2}\bigg)y(x)=E x^{N}y(x)\quad \ (N=-1,0,1,\dots).\eqno(1.3)$$
\nl Although the problem is posed here in one dimension, it often represents the radial equation for a problem in $d$-dimensions. For example, in the cases of Coulomb and harmonic oscillator problems, $m(m+1)/x^2$ might be the angular momentum term in $d$-dimensions with $m = \ell+(d-3)/2$. More generally, we might wish to consider generalizations of these potentials in which $m$ is real and non-negative.  Indeed, for the P\"oschl-Teller family of potentials, $m$ is simply a non-negative potential parameter.\medskip

The paper is organized as follows. In section 2, we review the asymptotic iteration method. This method was recently introduced\sref{\chs} to construct exact solutions for a wide class of Schr\"odinger equations and in many cases provided excellent approximate results for nontrivial eigenvalue problems\sref{\chs,\fern}. In section 3 we investigate the exact solutions of the basic eigenvalue problem (1.1). We obtain analytic formulae for the eigenvalues of (1.1) and also explicit expressions for the eigenfunctions. In section 4, we study the exact solutions of the differential equation (1.3) where the connection with the Bender {\it et al} class is discussed in detail. In section 6, we developed the exact solutions for different classes of P\"oschl-Teller Potentials. 
\title{2. The Asymptotic Iteration Method (AIM)}
\medskip
\noindent Consider the second-order homogenous linear differential equation
$$y^{\prime\prime}=\lambda_0(x)y^\prime+s_0(x)y\eqno(2.1)$$
for which  $\lambda_0(x)\neq 0$ and $s_0(x)$ are functions in $C_{\infty}(a,b)$. In order to find a general solution to this differential equation we rely on the symmetric structure of the right hand side of (2.1). Differentiating (2.1) with respect to $x$, we find that
$$y^{\prime\prime\prime}=\lambda_1(x)y^\prime+s_1(x)y\eqno(2.2)$$
where
$$ \lambda_1= \lambda_0^\prime+s_0+\lambda_0^2,\hbox{ and } s_1=s_0^\prime+s_0\lambda_0.$$
Meanwhile the second derivative of (2.1) yields
$$y^{\prime\prime\prime\prime}=\lambda_2(x)y^\prime+s_2(x)y\eqno(2.3)$$
for which
$$ \lambda_2= \lambda_1^\prime+s_1+\lambda_0\lambda_1,\hbox{ and } s_2=s_1^\prime+s_0\lambda_1.$$
Thus, for $(n+1)^{th}$ and $(n+2)^{th}$ derivative, $n=1,2,\dots$, we have
$$y^{(n+1)}=\lambda_{n-1}(x)y^\prime+s_{n-1}(x)y,\hbox{ and }\quad y^{(n+2)}=\lambda_{n}(x)y^\prime+s_{n}(x)y\eqno(2.4)$$
respectively, where
$$ \lambda_{n}= \lambda_{n-1}^\prime+s_{n-1}+\lambda_0\lambda_{n-1},\hbox{ and }  s_{n}=s_{n-1}^\prime+s_0\lambda_{n-1}.\eqno(2.5)$$
From the ratio of the $(n+2)^{th}$ and $(n+1)^{th}$ derivatives, we have
$${d\over dx}\ln(y^{(n+1)})={y^{(n+2)}\over y^{(n+1)}}=
{\lambda_n(y^\prime+{s_n\over \lambda_n}y)\over
\lambda_{n-1}(y^\prime+{s_{n-1}\over \lambda_{n-1}}y)}\eqno(2.6)$$
We now introduce the `asymptotic' aspect of the iteration method. For sufficiently large $n$, if
$${s_{n}\over \lambda_{n}}={s_{n-1}\over \lambda_{n-1}} := \alpha,\eqno(2.7)$$
then (2.6) reduces to
$${d\over dx}\ln(y^{(n+1)})=
{\lambda_n\over \lambda_{n-1}}\eqno(2.8)$$
which yields the exact solution
$$y^{(n+1)}(x)=
C_1\exp\bigg(\int\limits^x{\lambda_{n}(t)\over
\lambda_{n-1}(t)}dt\bigg) = C_1\lambda_{n-1}\exp\left(\int\limits^x(\alpha+\lambda_0)dt\right)\eqno(2.9)$$
where $C_1$ is the
integration constant, and the right-hand equation follows using (2.6) and (2.8).   Substituting (2.9) in (2.4) we obtain the
first-order differential equation
$$y^{\prime} + \alpha y = C_1\exp\left(\int\limits^x(\alpha+\lambda_0)dt\right)\eqno(2.10)$$
which, in turn, yields the general solution to (2.1)
as
$$y(x)= \exp\left(-\int\limits^{x}\alpha dt\right)\left[C_2 + C_1\int\limits^{x}\exp\left(\int\limits^{t}(\lambda_0(\tau) + 2\alpha(\tau)) d\tau \right)dt\right]
\eqno(2.11)$$
\title{ 3. An exactly solvable class of eigenvalue problem}
\medskip
\noindent In this section we use AIM to investigate the second order homogeneous linear differential equation
$$y^{\prime\prime}=2\left(a~x~p(x)-{{(m+1)}\over{x}}\right)y^\prime-wp(x)y,\eqno(3.1)$$
where $a$, $m$ and $w$ are real numbers. For special values of $a$ and $w$, this differential equation can immediately be integrated. For example, if $w=0$, then (3.1) yields
$$y(x)=C_1\int^x{e^{2a\int^tt'p(t')dt'}\over t^{2m+2}}dt+C_2.$$
\nl Furthermore, if $p(x)=e^{-{x^2/ w}}\bigg[{2(m+1)\over w}\int^x t^{-1}e^{at^2/w}dt+C_1\bigg],$
then
$$y(x)=e^{-\int^xwp(t)dt}\bigg[C_1\int^xe^{\int^t w p(t')dt'}dt+C_2\bigg].$$
\nl We are interested, however, in studying the exact solutions of (3.1) in which  
$p(x)={{x^N}\over{1-bx^{N+2}}},\ N=-1, 0, 1, 2, 3,...$,. In this case, (3.1) reads
$$y^{\prime\prime}=2\left({{ax^{N+1}}\over{1-bx^{N+2}}}-{{(m+1)}\over{x}}\right)y^\prime-{{wx^N}\over{1-bx^{N+2}}}y .\eqno(3.2)$$
Denote $$\lambda_0(x)=2\left({{ax^{N+1}}\over{1-bx^{N+2}}}-{{(m+1)}\over{x}}\right),\hbox{ and }\quad s_0(x)=-{{wx^N}\over{1-bx^{N+2}}},\eqno(3.3)$$
we may then apply AIM and the asymptotic condition (2.7) yields for $n=0,1,2,3,\dots$ 
\item{$\bullet$} $w_n^m(-1)=n(2a+2bm+(n+1)b)$ for $N=-1$
\item{$\bullet$} $w_n^m(0)=2n(2a+2bm+(2n+1)b)$ for $N=0$
\item{$\bullet$} $w_n^m(1)=3n(2a+2bm+(3n+1)b)$ for $N=1$
\item{$\bullet$} $w_n^m(2)=4n(2a+2bm+(4n+1)b)$ for $N=2$
\item{$\bullet$} $w_n^m(3)=5n(2a+2bm+(5n+1)b)$ for $N=3$
\item{$\bullet$} ... etc.

\noindent Thus, by induction on $N$, we can easily verify that the $w^m_n(N)$ are given by
$$w^m_n(N)=b~{(N+2)^2}~n~(n+\rho),\quad \rho={{(2m+1)b+2a}\over{(N+2)b}},\eqno(3.4)$$
where $N=-1,0,1,2,...$ and $n=0,1,2,3,...$.
For the exact solutions $y_n(x)$, we use the generator of the exact solutions (2.11), namely
$$y_n(x)= C_2~\exp\left(-\int\limits^{x}\alpha_k dt\right),\eqno(3.5)$$
where $n=0,1,2,...$ and $k \geq n$ is the iteration step number. Direct computations, using $\lambda_{k}= \lambda_{k-1}^\prime+s_{k-1}+\lambda_0\lambda_{k-1}$ and  $s_{k}=s_{k-1}^\prime+s_0\lambda_{k-1}$ where $s_0$ and $\lambda_0$ are given by (3.3),
imply the following:
\item{$\bullet$} $y_0(x)=1$ since $w_0^m(N)=0$ where $N=-1,0,1,2,....$.
\item{$\bullet$} $y_1(x)=-C_2(N+2)\sigma\left(1-{b(\rho+1)\over{\sigma}}x^{N+2}\right)$
\item{$\bullet$} $y_2(x)=C_2(N+2)^2\sigma(\sigma+1)\left(1-{{2b(\rho+2)}\over{\sigma}}x^{N+2}+{{b^{2}(\rho+2)(\rho+3)}\over{\sigma(\sigma+1)}}x^{2(N+2)}\right)$
\item{$\bullet$} $y_3(x)=-C_2{\sigma(\sigma+1)(\sigma+2)\over (N+2)^{-3}}\left(1-{3b(\rho+3)\over\sigma}x^{N+2}+{3b^2(\rho+3)(\rho+4)\over \sigma(\sigma+1)}x^{2(N+2)}-{b^3 (\rho+3)(\rho+4)(\rho+5)\over \sigma(\sigma+1)(\sigma+2)} x^{3(N+2)}\right)$
\item{$\bullet$} $\dots$ etc.

\ni Consequently, we arrive at the following general formula for the exact solutions $y_n(x)$
$$y_n(x)=(-1)^{n}{C_2(N+2)^{n}(\sigma)_n}~_2F_1\left(-n,\rho+n;\sigma;bx^{N+2}\right),\eqno(3.6)$$ 
where $(\sigma)_n={\Gamma(\sigma+n)\over\Gamma(\sigma)}$, $\sigma={{2m+N+3}\over{N+2}}$, and $\rho={{(2m+1)b+2a}\over{(N+2)b}}$. The Gauss hypergeometric function ${}_2F_1$ is defined by
$${}_2F_1(-n,b;c;z) = \sum\limits_{k=0}^n{(-n)_k(b)_k\over (c)_k}z^n,\eqno(3.7)$$
\nl that is to say, a polynomial of degree $n$ in $z$.
\title{4. A Class of exactly solvable eigenvalue problems and their generalization}
\medskip
\noindent An important step in solving eigenvalues problems using AIM\sref{\chs} is to find a suitable transformation that converts the eigenvalue problem under investigation into differential equation of the form (2.1) with $\lambda_0\neq 0$. As mentioned earlier (see\sref{\chs} and\sref{\fern}), the rate of convergence of AIM is influenced by this choice. The asymptotic solution of the eigenvalue problem usually provides a clear indication of a suitable transformation to use.  For the class of eigenvalue problems of Bender {\it et al} 
$$\left(-{{d^2}\over{dx^2}}+{{m(m+1)}\over{x^2}}+a^2x^{2N+2}\right)y(x)=Ex^{N}y(x), \quad 0\leq x< \infty,\eqno(4.1)$$
\nl where $N=-1,0,1,2,3,...$, $m\geq -1$, and  $y(x)$ satisfying the {\it Dirichlet boundary condition} $y(0)=0$, the asymptotic  solutions of the differential equation (4.1), for $x$ approaching 0 and $\infty$, suggest the following expression of the wave function: 
$$y(x)=x^{m+1}\exp\left(-{{ax^{N+2}}\over{N+2}}\right)f(x),\eqno(4.2)$$ 
where $f(x)$ is now to be determined by AIM.  Using (4.2) in (4.1), we find that the function $f(x)$ has to satisfy the differential equation
$$f^{\prime\prime}=2\left(ax^{N+1}-{{m+1}\over{x}}\right)f^\prime-wx^{N}f,\eqno(4.3)$$
where $w=E-a(2m+N+3)$. This differential equation is a special case of (3.2) with $b=0$ and $x\in[0,\infty)$. Thus (3.4) yields 
$$E=a(2n(N+2)+2m+N+3),\quad n=0,1,2....\eqno(4.4)$$ 
For the general solutions $f_n(x)$ ($n=0,1,2,\dots$), it is enough to take the limit in (3.6) as $b\rightarrow 0$. Thus the solutions of (4.2) can be written using the limit relation
$$\lim\limits_{b\rightarrow 0}{}_2F_1(-n,1/b+a;c;zb)={}_1F_1(-n;c;z).\eqno(4.5)$$
in the form
$$f_n(x)=(-1)^{n}C_2(N+2)^{n}(\sigma)_n~_1F_1\left(-n,\sigma;{{2a}\over{N+2}}x^{N+2}\right),\eqno(4.6)$$
\nl where $\sigma={{2m+N+3}\over{N+2}}.$

For $N=-1$, (4.1) represents a $d$-dimensional Schr\"odinger equation with the Coulomb potential, and the eigenvalues are $-a^2$, if $m$ is the angular momentum parameter $m = \ell + (d-3)/2$ in $d$-dimensions. Replacing the parameter $E$ (in this case $E$ becomes a coupling parameter) by $E=Ze^2$ and $a^2=-E_{nm}$, we easily recover, from (4.4), the well-known exact eigenvalues $E_{nm}=-{{Z^{2}e^{4}}\over{4(n+m+1)^2}}$ for the Coulomb problem in $d$ dimensions. Meanwhile, the case  $N=0$ represents the $d$-dimensional Schr\"odinger equation for the harmonic oscillator potential and the parameter $E$ itself is now the eigenvalue $E=E_{nm}$, where $m$ is again the angular momentum parameter $m = \ell + (d-3)/2.$ In this case, (4.4) yields the well-known exact eigenvalues  $E_{nm}=a(4n+2m+3)$ for this harmonic oscillator problem in $d$ dimensions. The parameter $m$ may be more generally any non-negative real number. For $N>0$ we emphasize that the eigenvalues of the boundary-value problem (4.1) do not correspond to those of a conventional Schr\"odinger equation. However, the zero eigenvalues may be the ground-state energies corresponding to an effective potential 
$$V(x)= {{m(m+1)}\over{x^2}}+a^2x^{2N+2}-gx^{N},\quad (g>0).$$ 
Indeed our method gives us the conditions on $m$, $a$, and $g$ under which such zero eigenvalues might occur. For example,
for $N=2$, we have for
$$\left(-{{d^2}\over{dx^2}}+{{m(m+1)}\over{x^2}}+a^2 x^6-gx^2\right)y(x)=Ey(x), \quad 0\leq x< \infty\eqno(4.7)$$
the energy is zero $E=0$ if
$$g=a(8n+2m+5),\quad n=0,1,2,\dots.\eqno(4.8)$$																
\title{5. P\"oschl-Teller Potentials}
\medskip
\noindent In this section we show how the asymptotic iteration method can be used to generate the exact solutions of the different families of P\"oschl-Teller potentials\sref{\pt-\lot}
$$V_{\rm I}(u)=k^2\left({\alpha(\alpha+1)\over \cos^2(k u)}+{\beta(\beta+1)\over \sin^2(ku)}\right)\quad\quad (0<k u<{\pi \over 2},~\alpha, \beta>0)\eqno(5.1)$$
and
$$V_{\rm II}(u)=k^2\left({\beta(\beta-1)\over\sinh^2(ku)}-{\alpha(\alpha+1)\over \cosh^2(ku)}\right)\quad\quad (0<ku<\infty, \alpha>\beta).\eqno(5.2)$$
Our approach shows that the exact solutions of P\"oschl-Teller potentials follow directly from solutions of the differential equation (3.2) through elementary transformations. The main idea is to use trigonometric or hyperbolic  mappings that make the potentials (5.1) and (5.2) rational and then make the corresponding Schr\"odinger equation suitable to be analyzed with AIM.
\title{5.1 The first class of P\"oschl-Teller potentials}
\medskip
We consider the one-dimensional Schr\"odinger equation of a quantum particle trapped by the P\"oschl-Teller potential $V_{\rm I}(u)$
$$-{{d^2y}\over{du^2}}+V_{\rm I}(u)y={\cal E}y\eqno(5.3)$$
where $V_{\rm I}(u)$ is given by (5.1). The potential $V_{\rm I}(u)$ is closely related to several other potentials which are widely used in molecular and solid state physics like the symmetric P\"oschl-Teller potential $\alpha=\beta \geq 0$, and the Scarf potential $-{1\over 2}\leq \alpha<0$.  The substitution $x=ku$, $0<x<{\pi\over 2}$, yields 
$$-{{d^2y}\over{dx^2}}+\left({{\alpha(\alpha+1)}\over{\cos^2(x)}}+{{\beta(\beta+1)}\over{\sin^2(x)}}\right)y=Ey,\quad (E={{\cal E}\over k^2})\eqno(5.4)$$
The potential $V_{\rm I}(x)={\alpha(\alpha+1)/ \cos^2(x)}+{\beta(\beta+1)/\sin^2(x)}$ is periodic. However, each period is separated from the next by an infinite potential barrier so it can be studied within one period, say $0< x< {\pi\over 2}$. Clearly, (5.4) is a smooth approximation for $\alpha,\beta\rightarrow 0^+$ of the infinite square-well potentials over the interval $[0,{\pi\over 2}]$. In order to apply the asymptotic iteration method, discussed earlier, to solve the eigenvalue problem (5.4), we have to transform it into a differential equation of the form (2.1) with suitable $\lambda_0\neq 0$. The asymptotic solutions of (5.4) as $x$ approaches $0^+$ and ${\pi\over 2}^-$ suggest for $y(x)$ the following expression 
$$y(x)=\cos^{\alpha+1}(x)\sin^{\beta+1}(x)f(x)\eqno(5.5)$$
where $f(x)$ is found by means of AIM. After substituting (5.5) into (5.4), we obtain the following differential equation for $f$
$$f^{\prime\prime}=2\left((\alpha+1)\tan(x)-(\beta+1)\cot(x)\right)f^\prime-wf\eqno(5.6)$$
where $w=E-(\alpha+\beta+2)^2$. Further, if we use the substitution $t=\cos(x),\ 0<t<1$, we arrive at the following equation
$$f^{\prime\prime}=2\left( {{(\beta+{{3}\over{2}})t}\over{1-t^2}}-{{\alpha+1}\over{t}}\right)f^\prime-{{w}\over{1-t^2}}f,\quad f^{\prime} = {{df}\over{dt}},\eqno(5.7)$$
which we may compare with (3.2) for $b=1$, $a=\beta+{{3}\over{2}}$, $m=\alpha$, and $N=0$. Thus from (3.4) we have 
$w=4n(\alpha+\beta+n+2)$ where $n=0,1,2...$. Finally, the eigenvalues for Eq.(5.4) are given by
$$E_n=(\alpha+\beta+2+2n)^{2},\quad n=0,1,2,\dots.\eqno(5.8)$$
Directly solutions $f_n(t)$ for the eigenvalue problem (5.8) can now be obtained from (3.6) with the substitution $b=1$ and $N=0$. Thus, we have
$$f_n(t)=(-2)^{n}C_2~(\sigma)_n~_2F_1(-n,\rho+n,\sigma;t^2),\quad t=\cos(x)\eqno(5.9)$$
where $\rho=\alpha+\beta+2$ and $\sigma=\alpha+{{3}\over{2}}$. Consequently, the wave functions for the first P\"oschl-Teller Potential (5.4) are
$$\eqalign{y_n(x)&=(-2)^{n}C_2(\alpha+{{3}\over{2}})_n~\cos^{\alpha+1}(x)\sin^{\beta+1}(x)~_2F_1(-n,\alpha+\beta+2+n,\alpha+{{3}\over{2}};\cos^2(x))\cr
&=(-2)^{n}C_2~n!~\cos^{\alpha+1}(x)\sin^{\beta+1}(x)~P_n^{(\alpha+{1\over 2},\beta+{1\over 2})}(1+2\cos^2x)}\eqno(5.10)
$$
using the relation\sref{\aar} between the hypergeometric function and Jacobi polynomial $P_n^{(\alpha,\beta)}(x)$  of degree $n$. The constant $C_2$ in (5.10) still to be determined by normalization.
\title{5.2. The second class of P\"oschl-Teller potentials}
\medskip
We now consider the Schr\"odinger equation 
$$-{d^2y\over{du^2}}+V_{\rm II}(u)y={\cal E} y,\quad 0<u<\infty \eqno(5.11)$$
where $V_{\rm II}(u)$ is given by (5.2). With the substitution, $x=ku$, we have
$$-{d^2y\over{dx^2}}+\left({{\beta(\beta-1)}\over{\sinh^2(x)}}-{{\alpha(\alpha+1)}\over{\cosh^2(x)}}\right)y=Ey,\quad 0<x<\infty \eqno(5.12)$$
We may assume $\alpha>\beta$, for if $\alpha<\beta$, we can change $\alpha\rightarrow -\alpha-1$, because the equation remains unchanged under $\alpha\rightarrow -\alpha-1$ and $\beta\rightarrow -\beta+1$. 
Because of the asymptotic solutions for $y$ as $x$ approaches $0^+$ and $\infty$, we may assume the exact form of the wavefunction $y(x)$ to be
$$y(x)=\cosh^{-\alpha}(x)\sinh^{\beta}(x)f(x),\eqno(5.13)$$
where the function $f(x)$ is to be found with AIM. After substituting (5.13) into (5.12), we obtain the following differential equation for $f(x)$
$$f^{\prime\prime}=2\left(\alpha\tanh(x)-\beta\coth(x)\right)f^\prime-wf,\eqno(5.14)$$
where $w=E+(\alpha-\beta)^2$. Further, by means of the substitution $t=\sinh(x)$, we can easily obtain
$$f^{\prime\prime}=2\left({{(\alpha-{{1}\over{2}})t}\over{1+t^2}}-{\beta\over t}\right)f^\prime-{{w}\over{1+t^2}}f\eqno(5.15)$$
where the prime refers to the derivative with respect to the variable $t$. This differential equation can be compared with (3.2) with $b=-1$, $a=\alpha-{{1}\over{2}}$, $m=\beta-1$ and $N=0$. Thus, using (3.4), we find the following expression for $w$:
$w=4n(\alpha-\beta-n)$, where $n=0,1,2...$. Finally, the eigenvalues for (5.12) are given explicitly by the expression
$$E_n=-(\alpha-\beta-2n)^{2},\quad n=0,1,2,\dots<{(\alpha-\beta)/2}.\eqno(5.16)$$
Consequently, the expected maximum number of quanta is
$$n_{max}=[(\alpha-\beta)/2]\eqno(5.17)$$
where $[(\alpha-\beta)/2]$ stands for the closest integer to $(\alpha-\beta)/2$ that is smaller than $(\alpha-\beta)/2$. 

With the substitution $b=-1$ and  $N=0$, we find using (3.6) that the exact solutions of $f_n(y)$ which satisfy (5.14) for $n=0,1,2,\dots<{(\alpha-\beta)/2}$  are
$$\eqalign{f_n(y)&=(-2)^{n}C_2~(\beta+{{1}\over{2}})_n~{}_2F_1(-n,\beta-\alpha+n;\beta+{{1}\over{2}};-t^2)\cr
&=(-2)^{n}C_2~n!~P_n^{(\alpha-{1\over 2},\beta-{1\over 2})}(1+2t^2)},\eqno(5.18)$$
where in the last line we have expressed the hypergeometric function in terms of the Jacobi polynomial in order to make easier the comparison with the results found in the literature\sref{\nie}. Up to a normalization constant the wave function $y(x)$ of (5.2) reads
$$y_n(x)=(-2)^{n}C_2(\beta+{{1}\over{2}})_n~\cosh^{-\alpha}(x)\sinh^{\beta}(x)~{}_2F_1(-n,\beta-\alpha+n;\beta+{{1}\over{2}};-\sinh^2(x)).\eqno(5.19)$$
\title{6.~~Conclusion}

In this paper we have applied the asymptotic iteration method to obtain the exact solutions for an interesting class of differential equations (3.2). We have shown that the exact solutions for the   class of problems studied by Bender {\it et al} are obtained by an application of AIM to a transformation of (3.2). This allows us to obtain the exact solutions for an extended class of eigenvalue problems (4.1) with $m\geq -1$.  Complete solutions for the well known coulomb and harmonic oscillator potentials follow directly by setting $N=-1,0$ respectively in (4.1). By means of trigonometric and hyperbolic coordinate transformations of $t$, in terms of which the first and second kind of P\"oschl-Teller potentials become rational functions of $t$, AIM also provides a direct way of generating the exact solutions to the Schr\"odinger eigenvalue problem generated by  these potentials. 

\title{Acknowledgments}
\medskip
\noindent Partial financial support of this work under Grant Nos. GP3438 and GP249507 from the 
Natural Sciences and Engineering Research Council of Canada is gratefully 
acknowledged by two of us (respectively [RLH] and [NS]).
\np

\references{1}

\end